\documentclass[prstab,reprint,showpacs,tightenlines, showkeys,superscriptaddress,amssymb,twocolumn]{revtex4-1}
\usepackage{amsmath,upgreek,graphicx,tikz}
\usepackage{subfigure}
\usepackage{color}
\usepackage{soul}
\usepackage{float}
\usepackage{lipsum}
\usepackage{comment}
\begin{document}

\title{PbTe(111) Sub-Thermionic Photocathode: A Route to High-Quality Electron Pulses}
\date{\today}

\author{Tuo Li}
\email{lizanchen1986@gmail.com}
\affiliation{Uptake Technologies, Inc., 600 W Chicago Ave Ste 620, Chicago, IL 60654, USA}
\author{W. Andreas Schroeder}
\affiliation{Department of Physics, University of Illinois at Chicago, 845 W. Taylor Street, Chicago,
IL 60607-7059, USA}

\begin{abstract} 
The emission properties of PbTe(111) single crystal have been extensively investigated to demonstrate that PbTe(111) is a promising low root mean square transverse momentum ($\Delta p_T$) and high brightness photocathode. The density functional theory (DFT) based photoemission analysis successfully elucidates that the `hole-like' $\Lambda_6^+$ energy band in the $L$ valley with low effective mass $m^*$ results in low $\Delta p_T$. Especially, as a 300K solid planar photocathode, Te terminated PbTe(111) single crystal is expected to be a potential 50K electron source.
\end{abstract}


\maketitle

\section{Introduction} 

The performance of Ultrafast Electron Microscopy (UEM)~\cite{PhysRevSTAB.10.034801, PhysRevLett.104.046801}, Dynamic Transmission Electron Microscopy (DTEM)~\cite{Lobastov17052005,:/content/aip/journal/apl/90/11/10.1063/1.2712838, PSSA:PSSA2210600215,:/content/aip/journal/rsi/74/10/10.1063/1.1611612}, Free Electron Lasers (FELs) and X-Ray Free Electron Lasers (XFELs) is fundamentally dependent upon the quality of photocathode generated electron beam in the transverse direction of the electron pulses produced by their front-end laser-driven electron gun. In the transverse spatial dimension, a high quality (high brightness) electron pulse requires a low normalized spatial transverse rms emittance. It is the aim of this paper to facilitate and demonstrate a dramatic improvement in the performance of spatial-resolved research instruments by investigating selected robust photocathode material exhibiting laser-driven emission with a low rms transverse momentum.\

A number of recent investigations, both experimental and theoretical, have aimed to gather insight into super-high brightness photocathodes~\cite{jin2008super, yamamoto2008high, nemeth2010high}. It is well known that the lead chalcogenides represent an important family of materials which have applications such as thermoelectric performance at high-temperature~\cite{ANIE:ANIE200900598}; conversion of heat and electricity~\cite{ANIE:ANIE200803934}, and the generation electricity from waste heat~\cite{snyder_complex_2008}. Among them, PbTe(111) is a well-studied narrow-gap semiconductor, as it can be employed both as an infrared detector~\cite{logothetis1971infrared, rogalski2010infrared} and a thermoelectric material~\cite{heremans2008enhancement, wang2011heavily}. The previous theoretical calculations of PbTe(111) in the rocksalt crystal structure implemented the quasiparticle self-consistent approach\cite{svane_quasiparticle_2010, PhysRevB.89.205203}, the augmented-plane-wave method\cite{rabii1968investigation} and $\vec{k}\cdot\vec{P}$ perturbation theory~\cite{PhysRev.135.A821}. These calculations are in good quantitative agreement with a large number of experiments~\cite{dalven1990introduction, leadsaltopticalandtransportmeasurements, PhysRevLett.6.667, PhysRev.173.714, cardona1964optical, stiles1961haas} investigating crystal structure properties, band structure, Fermi surface and effective mass. These previous studies shows that the valence and conduction band extrema occur at the $L$ point of the Brillouin zone (BZ), the `hole-like' $L_6^+$ valence band is relatively well isolated from other occupied bands at the $L$ high symmetry point. Both the conduction $L_6^-$ and valence band $L_6^+$ extrema have highly concentration-dependent masses. Importantly, PbTe(111) has a remarkable characteristic of its electric structure --- a low effective mass $\Lambda_6^+$ valence band in the vicinity of $L$ point; in particular, a very low effective mass (of the order of $10^{-2}m_0$) in the $L$--$K$ direction transverse to $\Gamma$--$L$ direction, which is perpendicular to the (111) face. According to the previous investigation of emission properties of metal photocathodes~\cite{bcc_jap, PhysRevSTAB.18.073401, PhysRevLett.111.237401}, low $m^*$ together with `hole-like' electron bands are preferred for low $\Delta p_T$, and rms transverse momentum for emission from such bands is almost electron temperature independent, therefore, PbTe(111) is an attractive candidate for low rms transverse momentum (high brightness) photocathode.\

In a bid to gain further insight to lead telluride for future high brightness electron source, in this paper, we report a benchmark study of the photoemission properties of such semiconductor, focusing particularly on the (111) crystal orientation relevant to low $\Delta p_T$ performance and its capability to generate high quality electron beam. In Sec. II, we confirm and supplement the electronic calculation of PbTe and outline a density functional theory (DFT) based analysis of photoemission. The summary discussion of the paper (Sec. III) directly explains the low $\Delta p_T$ of PbTe(111) as well as the acceleration field dependent $\Delta p_T$. Also we compare the experimental and the theoretical $\Delta p_T$ values of PbTe(111) with a cold atom electron source to show that the PbTe(111) single crystal is a high brightness photocathode --- a potential 50K electron emitter.\

\section{DFT-Based photoemission analysis}
The first step in the first-principles photoemission analysis~\cite{li2016photoelectric} is an evaluation of the electronic properties of PbTe(111). The calculations within DFT employ the PWscf code of the Quantum-EXPRESSO suite~\cite{QE-2009} and use ultrasoft pseudopotentials (USPPs) within the local density approximation (LDA)~\cite{garrity2014pseudopotentials}. Note that although GGA is more complex and indeed gives better bulk properties than LDA, it is well-known that LDA works better than GGA for certain classes of systems and properties, in particular for calculating the properties of many nonmetallic systems~\cite{mattsson2005designing}. This is because LDA shows a better cancellation of errors between surface exchange and correlation energies~\cite{hood1997quantum, hood1998exchange}. Therefore, LDA is selected as the exchange-correlation functional for the lead salt calculations. To carry out DFT calculations for a bulk crystal, the electronic wave function is described by plane-wave-basis sets with a kinetic energy cutoff of 28 Ry, and the energy cutoff for the charge density was set to 280 Ry. A threshold of $10^{-4}$ Ry/Bohr on the force for the ionic relaxation and 0.05 GPa on the pressure for the cell relaxation were used. A sampling of 6 $\times$ 6 $\times$ 6 Monkhorst-Pack~\cite{monkhorst_special_1976} set of special $k$-points and Marzari-Vanderbilt smearing~\cite{PhysRevLett.82.3296} with a broadening of 0.02 Ryd is employed. Full-relativistic effects are included in the DFT calculations, while spin-orbit coupling is included during the plane wave self-consistency iterations. Bulk PbTe has a rock-salt crystal structure with experimental low-temperature equilibrium lattice constant of 12.18 - 12.21 (a.u)~\cite{svane_quasiparticle_2010, madelung2005semiconductors, dornhaus1983narrow}. Each of the two atom types in the rock-salt structure forms a separate face-centered cubic lattice. The DFT calculated lattice constant is 12.01 (a.u), which is generally within 4.0$\%$ of the experimental values.\

\begin{figure}[htbp]
\vspace{-0.00cm}
\hspace{-0.00cm}
\center
\begin{tabular}{l l}
\includegraphics[scale=0.7]{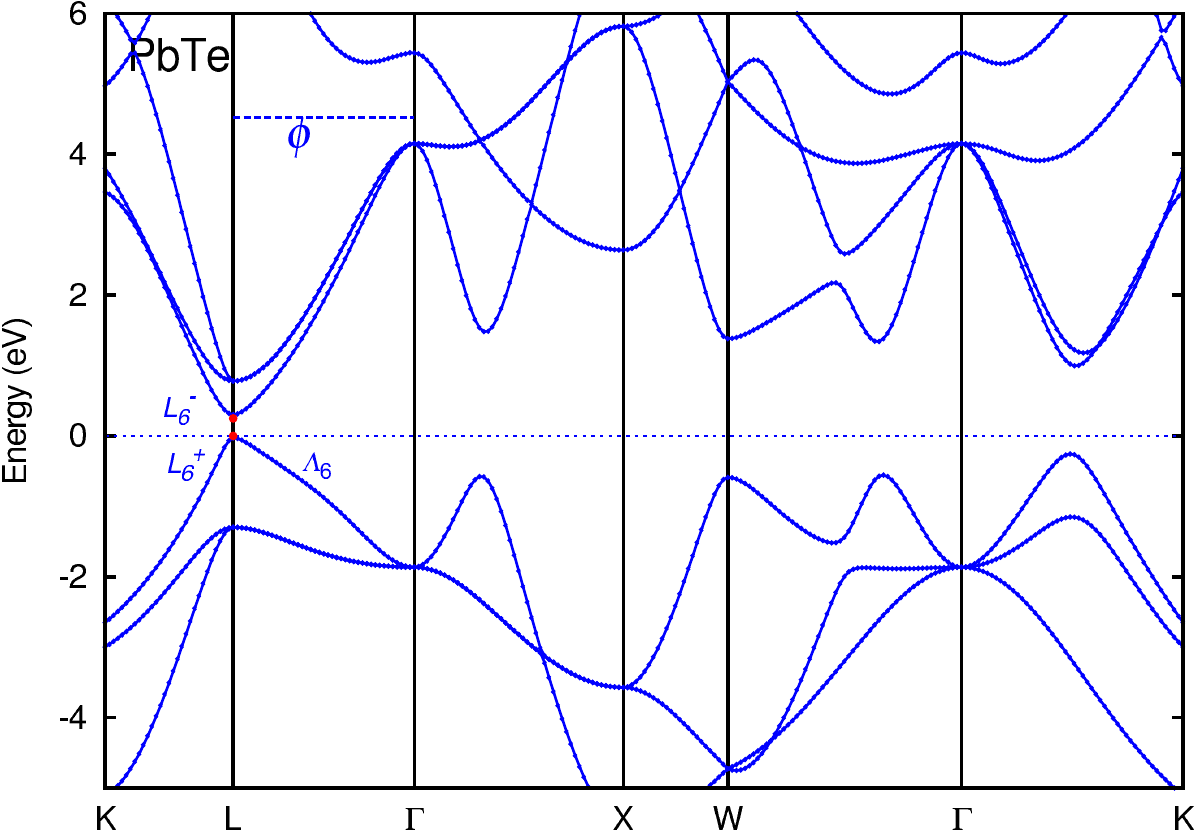}\\
\end{tabular}
\caption{PbTe band structure along major high-symmetry points and lines with $E_F$ (black dashed line) at zero energy placed  at the top of the VBM. The conduction band minimum (CBM) ($L_6^-$) and VBM ($L_6^+$) are labeled by red dots. $\Gamma \to L$ is the [111] direction which has the $\Lambda^+_6$ band with a low $m^*_T$ below the Fermi level. The calculated (111) face work function is indicated by a blue dashed line. The VBM state is an even $L^+_6$ state, while the CBM state is an odd $L^-_6$ state.}
\label{fig:PbXband.pdf}
\end{figure}
The band structure calculation uses the DFT evaluated lattice constant, the result is shown in Fig~\ref{fig:PbXband.pdf}. The states around the gap are dominated by Pb and Te highly hybridized $p$ bands~\cite{svane_quasiparticle_2010}, the average direct band gap is 2 - 3eV, while the minimum band gap occurs at the $L$ point with similar values of 0.26eV. The DFT calculated band structure of PbTe is generally within 5\% of the experimental low temperature values~\cite{pbtebandexpt}, especially, it shows a good agreement about the remarkable spiky $\Lambda_6^+$ valence band in the vicinity of $L$ point in both $L$ -- $K$ (transverse) and $\Gamma$ -- $L$ (longitudinal) directions. The experimental measured transverse effective mass ($m^*_T$)~\cite{dalven1990introduction} for the $\Lambda^+_6$ band (`hole-like') is 0.022$m_0$~\cite{dalven1990introduction}. It is important to note that the experimental measured transverse effective masses $m^*_T$ for the $\Lambda_6^+$ band of PbTe(111) is 0.022$m_0$ which is a factor of around 10 less than that of elemental metals~\cite{tsui_haas-van_1966, christensen_apw_1969, berlincourt_haas-van_1954, graebner_haas-van_1968, mattocks_haas-van_1978, baraff_haas-van_1973, joseph_haasvan_1964, windmiller_haas-van_1968, vuillemin_haas-van_1966, bhargava_haas-van_1967, surma_cyclotron_1979, dexter_cyclotron_1956, startsev_conduction_1984, hunt_haas-van_1989}. Due to the effect of effective mass on $\Delta p_T$~\cite{PhysRevLett.111.237401, bcc_jap, PhysRevSTAB.18.073401}, the $\Delta p_T$ obtained from metal
photocathodes is nearly the square root of one-third the product of excess energy and effective mass.\

The (111) surface of the PbTe could be terminated by either lead or chalcogen surface atoms. Although for the pure PbTe this surface is unreconstructed under all conditions, the chemical composition of the surface may be determined by measuring the intensity of the specularly reflected electron beam in reflection high-energy electron diffraction~\cite{frank1994imaging}, which clearly indicates that a surface dipole moment is formed naturally from alternatively terminated Pb and Te layers. For calculating the work function for the (111) cleavage face in PbTe, it is possible to build an `effectively charged' lead salt slab in three different ways based on the terminations of (111) cleaved surfaces: 1) Te terminated layers on both sides; 2) Pb terminated layers on both sides; 3) Te terminated and Pb terminated on one side each. These charged-slabs will result in different work functions due to the different dipole slab terminations. In order to simulate the relationship between work function and slab termination configuration, DFT work function calculations are performed on all three slab configurations. For all the slab configurations, the Fermi level is pinned just above the valence band maximum (VBM) by using 0.02Ry Gaussian spreading; therefore, the evaluated work function is defined as the energy offset from the VBM to the vacuum level. To ensure the accuracy of our work function calculations, the vacuum thickness is enlarged from 10 to 15$\AA$, the (1$\times$1) supercell thickness ($n$) is increased from 7 to 11 atomic layers together with $n\times n \times 1$ Monkhorst Pack points~\cite{PhysRevB.13.5188}, so the work function values converge within 0.05 eV. Dipole correction is added so that the contributions from electrostatic interaction due to the periodic boundary conditions is taken into consideration. As shown in Table~\ref{tab:Pbxwf}, the work function values increase as the slab termination changes from metal to chalcogen, which is consistent with the trend in the work function values from polycrystalline Pb (4.25eV) to Te (4.95eV)~\cite{michaelson_work_1977}. Khokhlov indicates that at elevated temperatures the lead salt (111) surfaces prefer to be metal terminated~\cite{khokhlov2002lead}, but all types of termination exist at room temperature. Therefore, the work function of PbTe(111) is as a function of the distribution of termination configurations, thus explaining the large ($\sim$ 0.5eV) variation in the experimentally measured values of $\phi_{(111)}$. Typical work functions of the three possible dipole configurations (Pb, Te, and Pb/Te) are generally less than 4.75eV thus allowing photoemission for 261nm ($\hbar\omega$ = 4.75eV) UV laser pulses obtained by harmonic conversion of a diode-pumped, 63MHz repetition rate, femtosecond Yb:KGW laser~\cite{berger2008high}. The thin-slab work function analysis indicates that $\phi_{(111)}$ ranges from 4.21 eV to 4.54eV for PbTe (See Table~\ref{tab:Pbxwf}) --- within $\pm$0.1 eV of the reported experimental range of 4.10 to 4.60eV. As a result, for incident 4.75eV (or even 4.67eV) UV photons, photoemission since the maximum possible transverse momentum $p_{T,max.}$ = $(2m_0\Delta E)^{1/2}$ of around 0.54$\sim$1.14 $(m_0\text{eV})^{1/2}$ should allow access to many electronic states in the PbTe(111) photocathodes. However, as shown below, the $\Lambda_6^+$ band structure (due to the low $m^*_T$) can play a major role in determining the maximum transverse momentum $p_{T,max}$ as well as $\Delta p_T$ of the photoemitted electrons.\
\begin{table}[ht!] 
\hspace{4.0cm}
\caption{PbTe(111) work functions in eV.} 
\begin{tabular}{l l l l l}  
\hline\hline
 & $\phi_{\text{Expt.}}$ & $\phi_{\text{DFT,Pb}}$ & $\phi_{\text{DFT,Te}}$ & $\phi_{\text{DFT,Pb/Te}}$ \\ [0.2ex]
\hline
PbTe(111) &4.10$^a$--4.60$^b$& 4.21 &4.54 &4.30\\ 
[1ex] 
\hline 
\end{tabular} 
\label{tab:Pbxwf}
\begin{tabular}{l}
$^a$Reference~\cite{nill1970laser}\\
$^b$Reference~\cite{bi_modeling_2009}\\
\end{tabular}
\end{table} 

\begin{figure}[!htbp]
\setlength{\tabcolsep}{20pt}
\centering
\vspace{0.00cm}
\hspace{0.00cm}
\begin{tabular}{@{}cc@{}}
\includegraphics[scale=0.22]{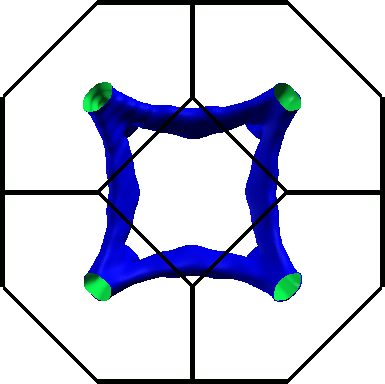}&
\includegraphics[scale=0.22]{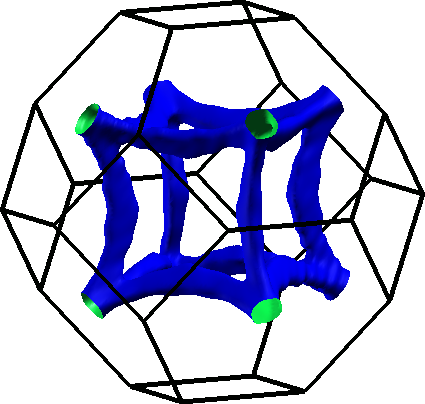}\\
\end{tabular}
\caption{The $E_F$-$\Delta E$ energy surface for PbTe with $\Delta E$=0.3 eV. The right panel is top view; the left panel is the (45$^o$, 45$^o$, 45$^o$) aerial view. The surface is isotropically distributed about the $\Gamma$--L axis.}
\label{fig:PbSfermi.png}
\end{figure}

The special features of PbTe in the interior of the first BZ can reveal the emission properties of PbTe in the transverse direction low $\Delta p_T$. As is well known, the Fermi level of semiconductor is pinned by its impurities --- the dopant. For undoped or weakly $p$-typed (111)-oriented PbTe single crystal, the Fermi level is just above the top of `hole-like' band dispersions at the $L$ point (VBM). Unlike elemental metals, that have electron bands across the Fermi level and available electron states at the Fermi energy, semiconductors do not have such an energy surface if the Fermi level is placed in the band gap. So, instead, Figure~\ref{fig:PbSfermi.png} shows the hole surface at a constant energy $E = E_{F}$ - 0.30 eV in the BZ for PbTe using a $20\times 20 \times 20$ uniform $k$-grid imposed on the 3D irreducible BZ. This is thus equivalent in photoemission with an excess energy $\Delta E$ = 0.3 eV, irrespective of crystal orientation and emission boundary conditions. It is clear that in [111] direction ($\Gamma \to L$) there is a small `muffin-tin' like surfaces (blue circle) with radii of 0.20$\sqrt{m_0eV}$ for PbTe.\

\begin{figure}[h]
\centering
\vspace{-0.00cm}
\hspace{-0.00cm}
  \begin{tabular}{@{}cc@{}}
\includegraphics[scale=0.32]{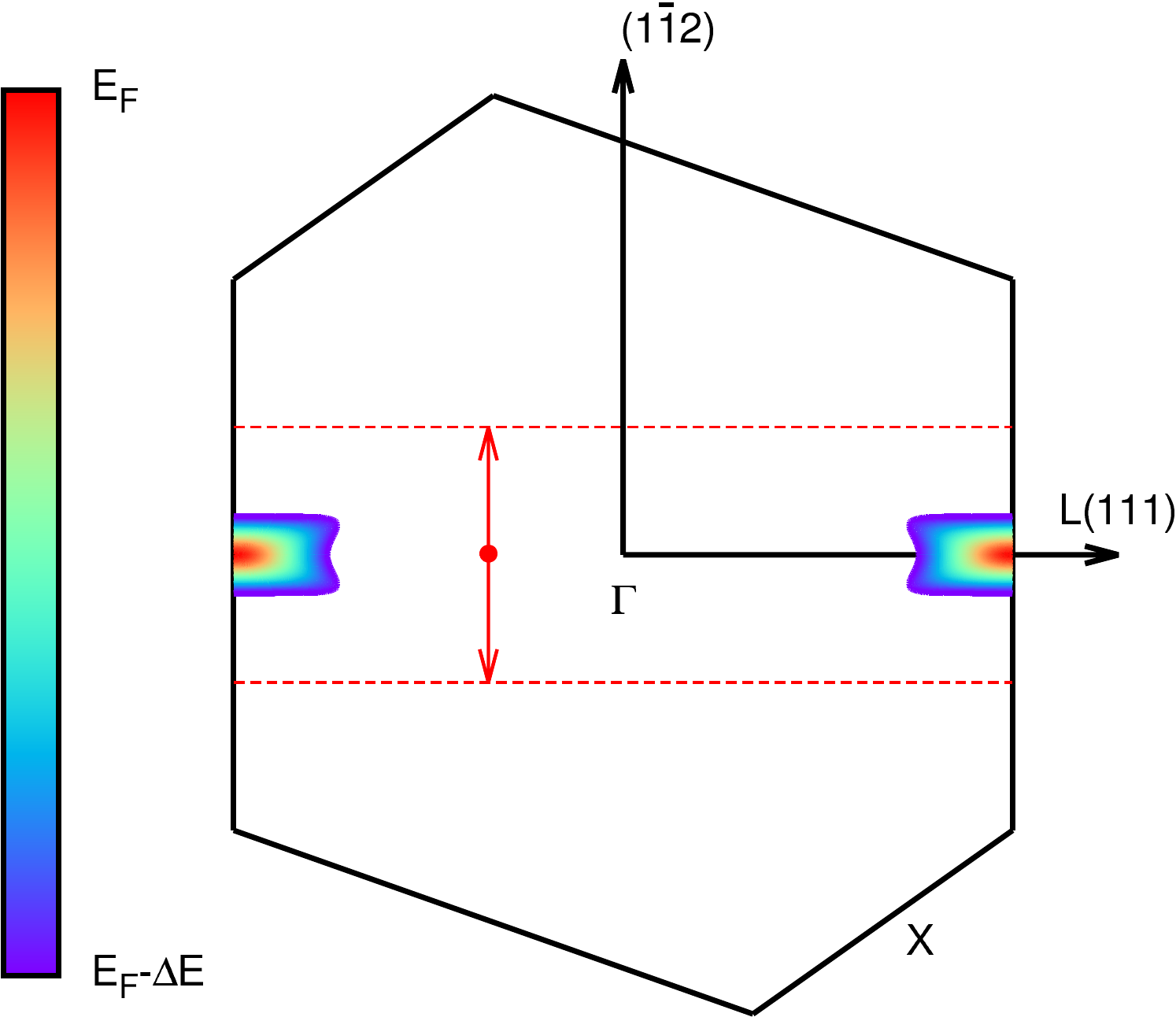}&
\includegraphics[scale=0.64]{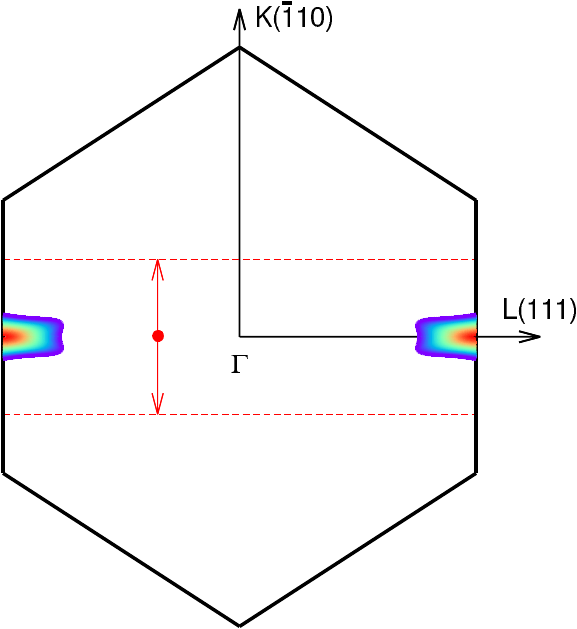}\\
(a) & (b) \\
\end{tabular}
\caption{PbTe(111) face photoemitting electron energy states contour for excess energy $\Delta E$ = 0.3 eV. The $p_{T,max}$ values are labeled by the red dashed lines. The color palette from red to blue indicates the energy level from $E_F$ to $E_F-\Delta E$. (a) The (1$\overline{1}$2) transverse momentum face. (b) The (1$\overline{1}$0) transverse momentum face.}
\label{fig:PbTe_111_110contour.pdf}
\end{figure}

Figure~\ref{fig:PbTe_111_110contour.pdf} shows the PbTe(111) face momentum distributions in the transverse (1$\overline{1}$2) and (1$\overline{1}$0) directions where $\Delta E$ = 0.3eV. The photoemitting contours are condensed and located at the edge of $L$ point with a `muffin-tin' shape in the first BZ. For $\Delta E$ = 0.3eV, the theoretical $\Delta p_{T,max}$ equals 0.775$\sqrt{m_0eV}$, which is indicated by the red dashed lines. Clearly, the `hole-like' $\Lambda_6^+$ valence band itself plays the dominant role in confining the maximum transverse momentum within the band tip so that the transverse momentum does not spread from 0 to 0.775 $\sqrt{m_0eV}$. In fact, the low values of $m^*_T$ for the $\Lambda_6^+$ band limits $p_T$ to less than half of $p_{T,max} = \sqrt{2m_0(\Delta E)}$ --- a band bending constraint on $p_T$ that is not present in most metals and their alloys~\cite{bcc_jap, PhysRevSTAB.18.073401}. This important characteristic of PbTe(111) photocathodes will lead to the emission of well converged (low $\Delta p_T$) electron beams.\

\section{Results and Discussion}
\begin{figure}[ht!]
\centering
\vspace{-0.00cm}
\hspace{-0.00cm}
\begin{tabular}{@{}cc@{}}
\includegraphics[scale=0.25]{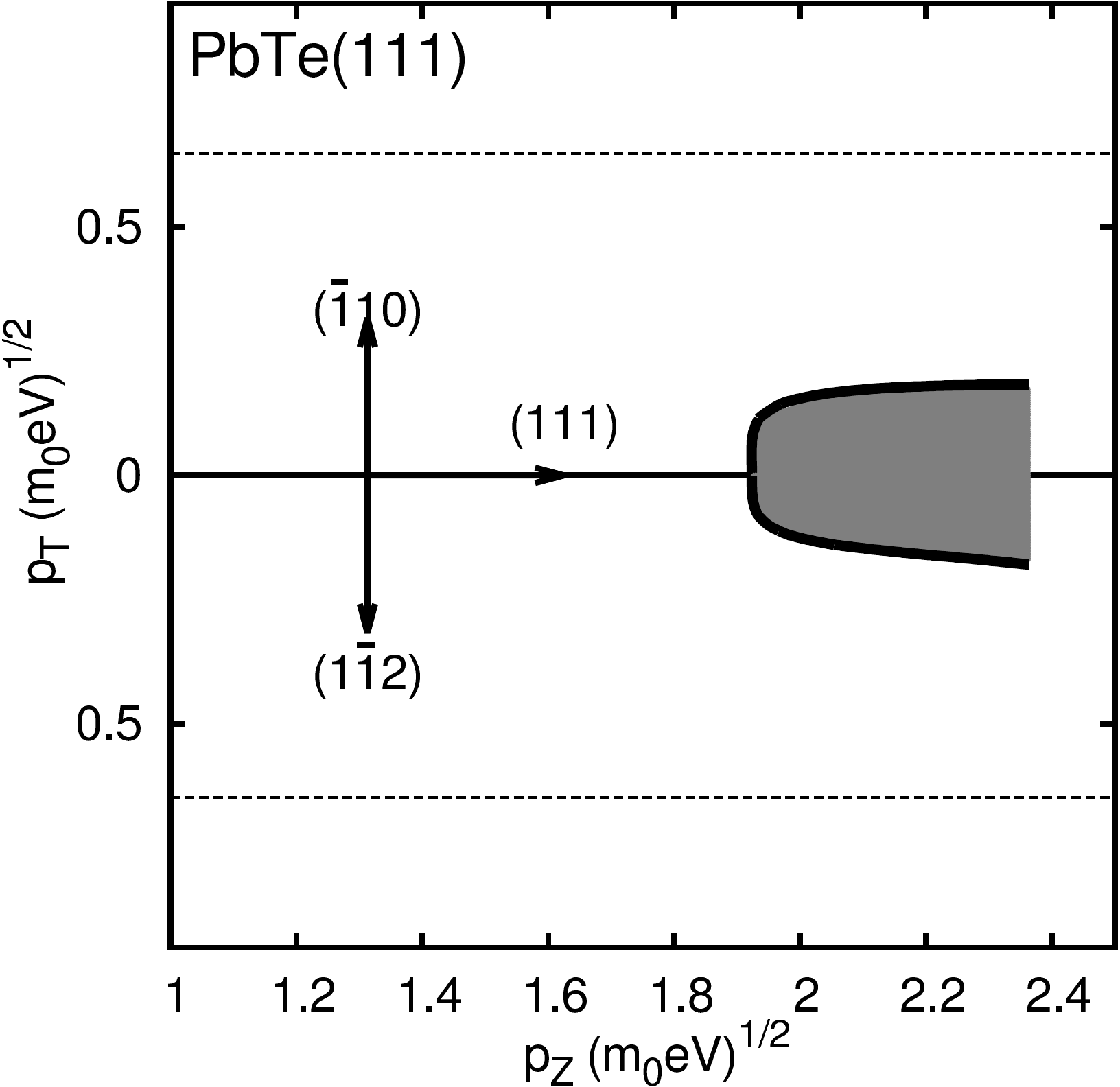}&
\includegraphics[scale=0.25]{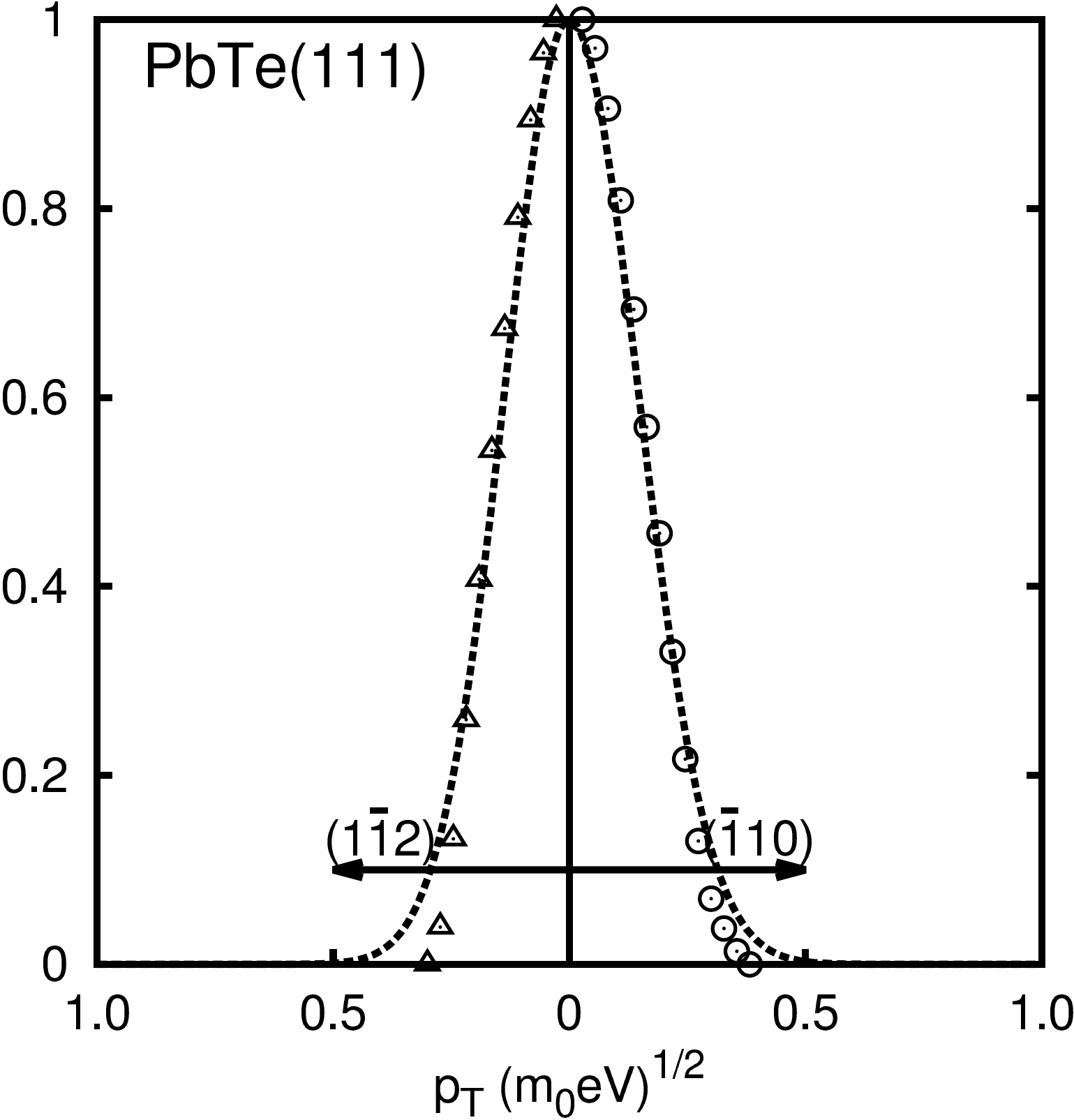}\\
(a) & (b) \\
\end{tabular}
\caption{Te terminated PbTe(111) photoemission states and transverse momentum distribution under $\hbar\omega$ = 4.75eV: (a) Crystal momentum map of the electronic states (shaded regions) 0.21eV below the Fermi level (solid line) that may photoemit within the real $p_{T,max}$ (dashed line) for the transverse (1$\overline{1}$2) and (1$\overline{1}$0) crystal directions; (b) Transverse momentum distributions of the photoemitted electrons in the (1$\overline{1}$2) and (1$\overline{1}$0) directions (Gaussian fits are guides to the eye).}
\label{fig:PbTept.pdf}
\end{figure}

The results from the DFT-based analysis for photoemission from the (111) face of PbTe with $\hbar\omega$ = 4.75eV using the work function calculated for the Te terminated slab, i.e, 4.54 eV is displayed in Fig.~\ref{fig:PbTept.pdf}. The crystal momentum depiction of the photoemitting electronic states (shaded regions) $\Delta E$ = 0.21eV below the Fermi level (solid line) in Fig.~\ref{fig:PbTept.pdf} (a) clearly shows that emission is from the `hole-like' states associated with the $\Lambda^+_6$ band Fermi surface centered on the $\Gamma$-point of the Brillouin zone. The emitting states are very isotropic about the primary [111] direction and have $p_T$ maximum of 0.21$\sqrt{m_0eV}$, which is a factor of nearly 3 less than the value expected from $p_{T,max} = \sqrt{2m_0(\Delta E)}$. As a result, the transverse momentum distribution for the emitted electrons (See Fig.~\ref{fig:PbTept.pdf} (b)) is quite narrow. In addition, the PbTe band structure calculation (See Fig.~\ref{fig:PbXband.pdf}) indicates that there are no upper conduction band minimum within in 3eV of the vacuum energy --- thus eliminating the possible of excited states thermionic emission with higher $p_T$~\cite{berger_excited-state_2012}. For the work functions associated with the three termination configurations (See Table~\ref{tab:Pbxwf}), the calculated $\Delta p_T$ of PbTe(111) is 0.067$\sqrt{m_0eV}$, 0.125$\sqrt{m_0eV}$, and 0.152$\sqrt{m_0eV}$ for the Te, Te/Pb, and Pb terminated slabs, respectively. As expected, with the same incident photon energy, a Te terminated PbTe(111) single crystal photocathode is predicted to have the lowest $\Delta p_T$ as it also has the smallest $\Delta E$.
\

\begin{figure}[h]
\centering
\begin{tabular}{@{}c@{}}
\includegraphics[scale=0.44]{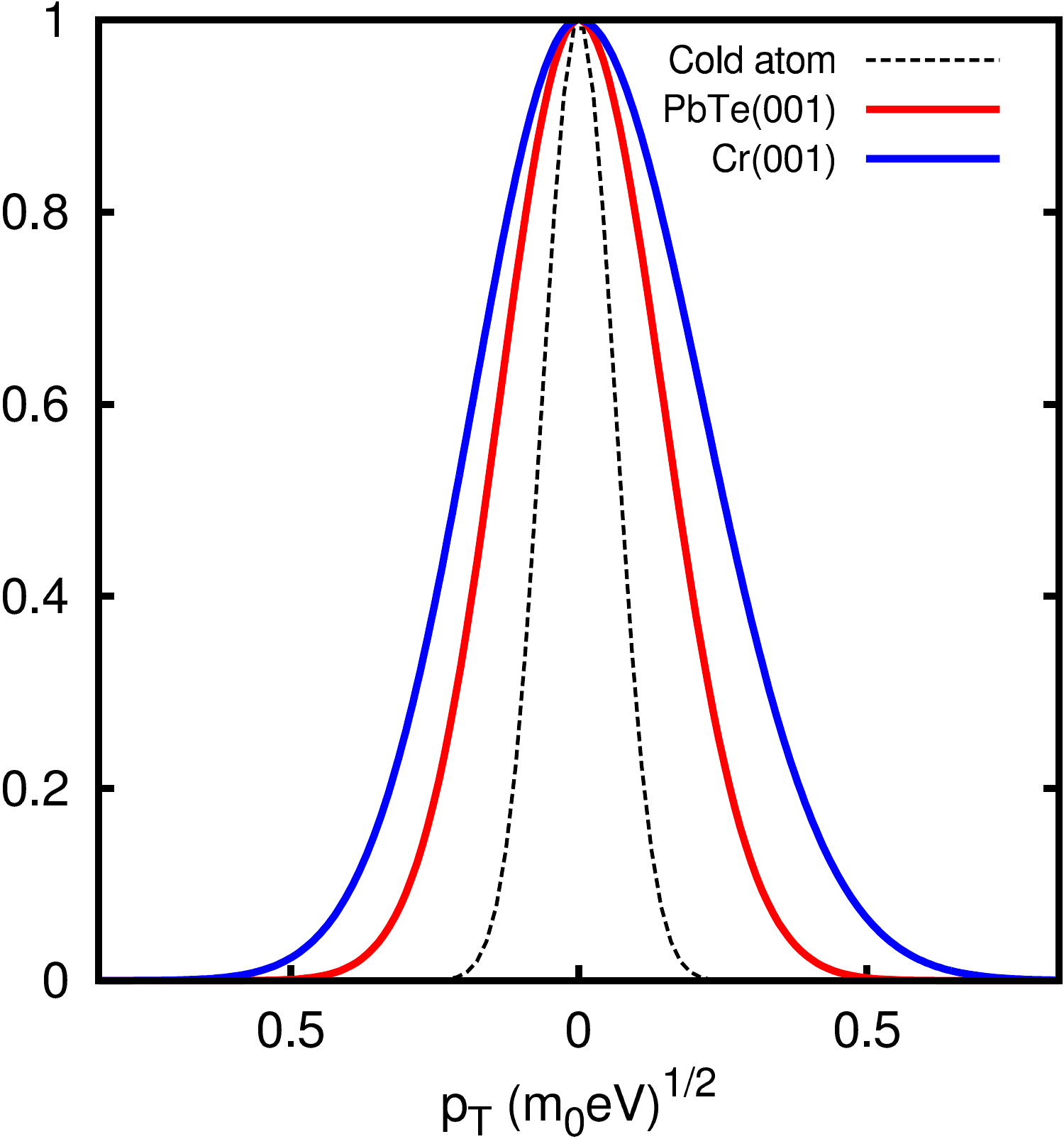}\\
\end{tabular}
\caption{(a) Comparison of transverse momentum distributions among the PbTe(111), Cr(001) face and the cold atom source of Ref.~\cite{Saliba:12}.}
\label{fig:leadpT.pdf}
\end{figure}

\begin{figure}[ht!]
\centering
\vspace{-0.00cm}
\hspace{-0.00cm}
\begin{tabular}{@{}c@{}}
\includegraphics[scale=0.5]{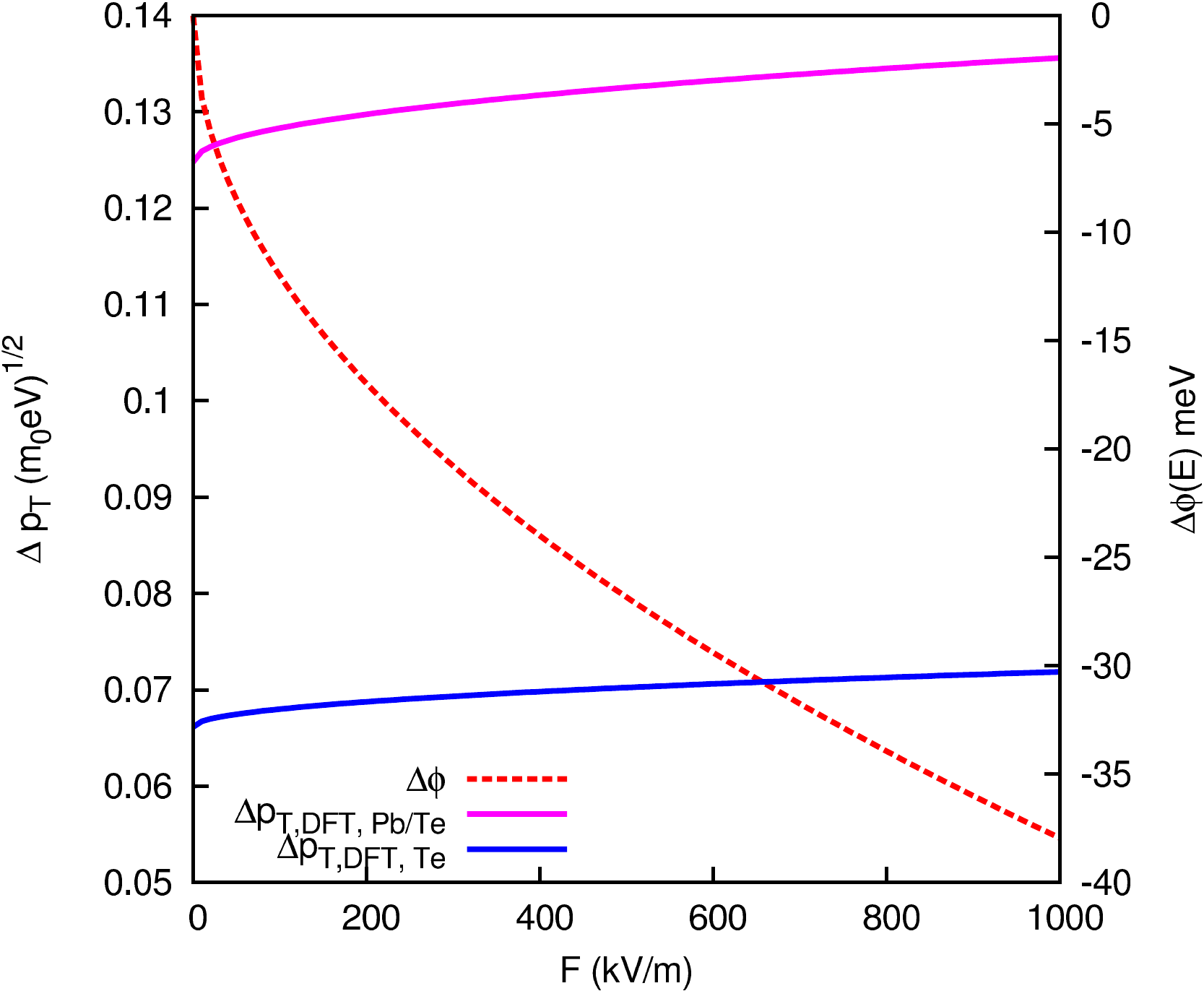}\\
\end{tabular}
\caption{Schottky effect on $\Delta p_T$. The relationship between acceleration field $\Delta \phi = \sqrt{e^3F/4\pi\epsilon_0}$ is in red dashed line. For Te (Pb/Te) terminated PbTe(111), the $\Delta p_{T, \text{DFT}}$ as a function of $F$ is shown in blue (pink) solid line.}
\label{fig:schottky.pdf}
\end{figure}

The expected transverse momentum distribution of a cold atom photoelectron source at 14($\pm$2)K~\cite{Saliba:12} is shown in Fig.~\ref{fig:leadpT.pdf} (black dashed line), assuming a Gaussian distribution of the form exp$[-\frac{p^2_T}{2\Delta p^2_T}]$ with $\Delta p_T$ evaluated using $\sqrt{m_0k_BT}$ = 0.0347$\pm$0.0131$\sqrt{m_0eV}$. For comparison, the transverse momentum distributions of Te terminated PbTe(111) and Cr(001)~\cite{bcc_jap} are also displayed in Fig.~\ref{fig:leadpT.pdf} with the solid red and blue lines, respectively. For a 300K solid planar photocathode, PbTe(111)'s predicted $\Delta p_T$ is only a factor of 2 more than that of the cold atom source, and a factor of 1/2 less than that of Cr(001). According to the expression for $\Delta p_T$ for thermionic electron emission $\sqrt{m_0k_BT}$~\cite{PhysRevSTAB.12.074201}, Te terminated PbTe(111) is close to a 50K electron source. It is important to note that the DFT evaluated $\Delta p_T$ for PbTe(111) is at zero temperature. So the experimental measured value may be above this value due to temperature effects (e,g. thermal excitation across the band gap), sample termination configuration, and other factors (e.g., oxidation and Schottky effect) which can affect the effective work function. Especially, the $\Delta p_T$ is also a function of acceleration field due to the Schottky effect~\cite{simmons1967poole} caused work function change, which can be expressed by $\Delta \phi = \sqrt{e^3F/4\pi\epsilon_0}$, where $\epsilon_0$, $F$, $e$ are vacuum permitivity, electric field magnitude, and electron charge, respectively. As the work function is decreasing with respect to the acceleration field $F$, the $\Delta p_T$ is monotonically increasing when the magnitude of acceleration field becomes greater. Figure~\ref{fig:schottky.pdf} shows that the Schottky effect caused work function change is of the order of 10$^{-2}$ eV when $F$ varies from 0 MV/m to 1MV/m, so the change of $\Delta p_T$ is within 4\% of the $\Delta p_{T,F=0}$.

It is, however, notable that the transmission flux probability, $T(p_z, p_{z0})$, over the work function barrier is significantly larger for PbTe(111) emission than most metals, where $p_z$ ($p_{z0}$) is the longitudinal momentum of electron inside (outside) the photocathode~\cite{bcc_jap}. This is because the combination of a relatively small longitudinal effective mass $m_z^*$ and the additional crystal momentum at the $L$ point VBM allows emission close to $p_{z0}$ = $p_z$ where $T(p_z, p_z0)$ is maximized. On the other hand, the density of possible excited states is less in PbTe than most metals due to the much smaller effective mass of the VB in PbTe. The net result is that the quantum efficiency of PbTe is expected to be comparable to that of a metal.\

In practice, the brightness of a next-generation cold electron unstructured photoemission source cannot arbitrarily increase as $\Delta p_T$ is reduced. J. M. Maxson has already anticipated such a limit to be approached in the next generation of high brightness electron sources producing intense beams~\cite{maxson2013fundamental}; namely, the thermally induced disorder~\cite{maxson2013fundamental} limits in which any `cold' electron gas quickly equilibrates its temperature so that the kinetic energy associated with the rms momenta of electrostatic the electron state potential energy associated with their rms separation. For a $\sim$50K PbTe(111) photocathode, this could limit the density of the generated electron pulse to $10^{19}m^{-3}$. There are numerous solid-state compounds that could be used as photocathodes, this work has outlined an essential requirement for an ultra-low $\Delta p_T$ planar solid-state photoemitter; namely, ideally a low $m^*$ `hole-like' band, the Te terminated PbTe(111) single crystal is an example of this approach to new photocathode discovery. The computational complexity in determining accurate work function values for solid-state compounds results from surface terminations (as shown in Table~\ref{tab:Pbxwf}), oxidation, etc.) means that a more direct connection of the DFT-based photoemission simulation to experiment will require $situ$ measurement~\cite{zhang1995performance} of the photocathode work function during experiments (such as the solenoid scan technique~\cite{PhysRevLett.111.237401}) aimed at determining $\Delta p_T$. Therefore, a tunable UV laser radiation source will be needed to realize such exciting avenue for future research.\
\begin{acknowledgments}
The authors are indebted to Juan Carlos Campuzano, Christopher Grein, Randall Meyer, Alan Nicholls, and Serdar \"{O}\u{g}\"ut for their valuable discussions. This work was partially supported by the Department of Energy (Contract No. DE-FG52-09NA29451).
\end{acknowledgments}
\bibliography{PhD.bib}
\end{document}